\documentclass[amsmath,amssymb,12pt,a4paper,nofootinbib, showpacs,showkeys]{revtex4-1}
\usepackage{amssymb,amsmath}

\usepackage{graphicx}

\newcommand{\bomega}{\mbox{\boldmath{$\omega$}}}

\newcommand{\R}{\mathbb{R}}
\newcommand{\sref}[1]{{Section \ref{#1}}}
\newcommand{\eref}[1]{{(\ref{#1})}}
\newcommand{\mbf}[1]{{\mathbf{#1}}}
\begin{document}

\title {3D Euler equations and ideal MHD mapped to regular systems:\\ probing the finite-time blowup hypothesis}

%\vspace{-.4cm}

\author{Miguel D. Bustamante}

\address{School of Mathematical Sciences, University College Dublin, Belfield, Dublin 4, Ireland, EU}

\email{miguel.bustamante@ucd.ie}

%\vspace{-.6cm}

\begin{abstract}
We prove by an explicit construction that solutions to incompressible 3D Euler equations defined in the periodic cube
$\Omega = [0,L]^3$ can be mapped bijectively to a new system of equations whose solutions are
globally regular. We establish that the usual Beale-Kato-Majda criterion for finite-time singularity (or blowup) of a
solution to the 3D Euler system is equivalent to a condition on the corresponding \emph{regular} solution of the new
system. In the hypothetical case of Euler finite-time singularity, we provide an explicit formula for the blowup time in
terms of the regular solution of the new system. The new system is amenable to being integrated numerically using similar
methods as in Euler equations. We propose a method to simulate numerically the new regular system and describe how to use
this to draw robust and reliable conclusions on the finite-time singularity problem of Euler equations, based on the
conservation of quantities directly related to energy and circulation. The method of mapping to a regular system can be extended to any fluid equation that admits a Beale-Kato-Majda type of theorem, e.g. 3D Navier-Stokes, 2D and 3D magnetohydrodynamics, and 1D inviscid Burgers. We discuss briefly the case of 2D ideal magnetohydrodynamics. In order to illustrate the usefulness of the mapping, we provide a thorough comparison of the analytical solution versus the numerical solution in the case of 1D inviscid Burgers equation.

\end{abstract}



\maketitle

\newpage

\section{Introduction}
One of the most important unsolved problems in Mathematics entails a simple question: are solutions to the
three-dimensional Euler equations globally regular or do they blow up in a finite time? The analogous question for
Navier-Stokes equations, also unsolved, corresponds to one of the famous Millennium Prize problems \cite{Millennium}. The
analytical and numerical methods appearing in the scientific literature to solve these equations have been highly
transferrable to real-life problems, such as high-Reynolds number turbulence and vortex reconnection in Navier-Stokes,
magnetic reconnection in magnetohydrodynamics, and extreme events in the atmosphere, to mention a few.

While several conclusive results are available for Euler and Navier-Stokes in two dimensions, three dimensions with axial
symmetry and three dimensions with helical symmetry
\cite{Yudovich63,Kat67,Ladyzenskaja68,Ukhovskii68,Mahalov90,Dutrifoy99,Ettinger09}, attempts to understand the regularity of
three-dimensional solutions have only reached local and/or conditional analytical results
\cite{Ebi70,Kat72,Bou74,Tem75,BKM84,Pon85,Fer93,Cons94,Consetal96,KozTan00,Cor01,Dengetal03,Cha03,Cha04,Cha05,Dengetal05,Dengetal05a,Chae07,Gib07,Gib08}. Since the advent of powerful computers circa 1980, numerical researchers have also contributed with competing
yes/no conclusions regarding finite-time singularity, mainly in three-dimensional Euler, which is the focus of the present
paper \cite{Bar76,Morf80,Cho82,Bra83,Sig84,Kid85,Pum87,Ash87,Pum90,Bel92,Bra92,Ker93,Bor94,Pel97,Gra98,Pel01,Ker05a,
 Pel05,Gul05,Cich05,Ker05,Pau06,HouLi06,Orl07,Graf08,HouLi08,BusKerr08}.

 In a 3D Euler numerical simulation the spatial distribution of vorticity tends to get localised in structures that become
increasingly sharp with time. This entails a finite-time loss of resolution, not necessarily due to a true finite-time
singularity of the solutions, but rather to a finite amount of memory available for a computer simulation. One can identify
two main drawbacks in current and previous state-of-the-art numerical attempts on the Euler singularity problem. On the one
hand, it is not known analytically whether a given initial condition can give rise to a finite-time singularity, hence a
numerical solution of 3D Euler equations is inherently ``blind'' to any potential finite-time singularity that may be
encountered: numerical dissipation and loss of resolution may certainly ``shield'' the singularity. On the other hand,
instability and numerical error could give rise to a ``fake'' finite-time singularity, typically associated with a lack of
resolution at late stages of a simulation.

This observation motivated the current research: our main analytical result is that there is a bijective mapping from Euler
equations (along with velocity fields) to a new system of equations, hereby called \emph{mapped}, whose solutions are
globally regular. The mapped system is amenable to direct numerical integration using the same methods as in Euler
equations, with the important advantage that the solutions of the mapped system are regular by definition. In this way, and
for the first time ever, reliable and robust conclusions on the problem of finite-time singularity of Euler equations could
be drawn, by analysing in post-processing the data from a careful numerical integration of the mapped system.

In Section \ref{sec:Euler} we define the bijective mapping from 3D Euler to a regular system, provide the proofs of global regularity, and show how to use the solution of the regular system to probe the hypothesis of finite-time singularity of the 3D Euler equations. In Section \ref{sec:proposed} we propose a numerical method to implement these ideas and discuss the advantages and potential challenges of the new approach. In Section \ref{sec:other} we show that other equations of interest in physics and mathematics can be mapped to regular systems, provided they admit a Beale-Kato-Majda (BKM) type of theorem (e.g., Navier-Stokes, magnetohydrodynamics and inviscid Burgers), and we construct an explicit mapping from the 2D ideal magnetohydrodynamics equations to a regular system. In Section \ref{sec:Burgers} we illustrate the usefulness of the mapping to a regular system, by considering a simple equation admitting a BKM type of theorem: the 1D inviscid Burgers equation. We compare thoroughly the analytical solution for the supremum norm of velocity gradient, with the numerical data from both direct numerical simulation of inviscid Burgers and numerical simulation of the mapped system, and conclude that the integration of the regular mapped system is far more accurate and converges better than the original system. Finally, we present concluding remarks in Section \ref{sec:concl}.

\section{Euler equations: previous vs. novel approach}
\label{sec:Euler}
%In this section we introduce notation for the 3D Euler equations in a periodic cube, describe the main assumptions on the
%fields and motivate the introduction of the mapping.

\subsection{Euler equations in a periodic cuboid}

The Euler fluid equations describe the evolution of an incompressible, unit mass density flow with velocity field
$\mbf{u}(x,y,z,t) \in \R^3$ defined for $(x,y,z) \in \R^3$ and in a time interval $t \in [0,T)$:
\begin{eqnarray}
\label{eq:Euler}
 \frac{\partial \mbf{u}}{\partial t} + \mbf{u} \cdot \nabla \mbf{u} = - \nabla p\,, \qquad \quad \nabla \cdot \mbf{u} = 0.
\end{eqnarray}
Periodic boundary conditions are assumed for the velocity field and pressure with a basic periodicity domain $ \Omega =
[0,L]^3$. That is, for any $(x,y,z) \in \R^3$ we have $ \mbf{u}(x+L,y,z,t) = \mbf{u}(x,y+L,z,t) = \mbf{u}(x,y,z+L,t) =
\mbf{u}(x,y,z,t), \,\forall\,t \in [0,T)$ and similarly for $p$. The basic periodicity domain is by definition the smallest
domain with such periodicity property.

Here, and throughout this proposal, $T$ denotes a generic time so that $\mbf{u} \in C([0,T);H^s)\cap
C^1([0,T);H^{s-1})\,,\,\,s\geq3\,,$ so in particular the Sobolev norms of the velocity field are bounded: 
\begin{equation}
 \label{eq:Sobolev}
{{||\mbf{u}(\cdot,t)||_{H^s} \equiv \left({\displaystyle\sum\limits_{\mbf{k}\in \mathbb{Z}^3}} (1 +
|\mbf{k}|^2)^s\,|\widehat{\mbf{u}}(\mbf{k},t)|^2\right)^{\frac{1}{2}} \leq c_s\,,\,\,t\in[0,T)\,,\,\,s\geq3\,,}}
\end{equation}
 where $c_s$ are some constants and $\widehat{\mbf{u}}(\mbf{k},t)$ are the Fourier coefficients of the velocity field.

\subsection{Previous analytical and numerical results: vorticity and the finite-time singularity hypothesis}

In order to gain some insight on the problem of finite-time
blowup, the most common approach is to monitor from numerical simulations a number of quantities pertaining to the theory.
The main quantity studied since early numerical simulations (ca.~1987) is the vorticity supremum norm
$||\mathbf{\bomega}(\cdot,t)||_{\infty} \equiv {\sup_{\mbf{x}\in \Omega }} |\bomega(\mbf{x},t)|\,,$ where $\bomega \equiv
\nabla \times \mbf{u}$ is the vorticity field. According to a theorem by Beale, Kato and Majda (BKM) \cite{BKM84}, the
assumed regularity of the velocity field can be extended up to and including the time $T$ \mbox{if and only if}
\mbox{$\tau(T) \equiv \int_{0}^T ||\bomega(\cdot,t)||_{\infty} ~dt < \infty.$} By `regular up to and including the time
$T$' we mean $\mbf{u} \in C([0,T];H^s)\cap C^1([0,T];H^{s-1}),\,s\geq3$.

\emph{The hypothesis of finite-time singularity (blowup)} states that there is a `singularity time' $T \in (0,\infty)$,
such that $\tau(T) = \infty.$ It is an open problem to establish analytically the validity of this hypothesis.
Consequently, it has been an important aim of many numerical simulations of Euler equations \eref{eq:Euler}, published
since BKM theorem emerged, to conclude \emph{yes}, \emph{no} or \emph{maybe} on the validity of this hypothesis.
Unfortunately, direct numerical simulations of Euler equations 
entail an inherent uncertainty about the time of a hypothetical singularity. Therefore, current numerical attempts on the
Euler singularity problem cannot give 100\% reliable conclusions in favour of or against a finite-time singularity. A fresh
approach is urgently needed to continue along this research path.

\subsection{The novel approach}
\label{subsec:novel}
The main goal of this paper is to provide such a fresh approach to study the
validity of the hypothesis of finite-time singularity. This is based on a bijective mapping which maps the Euler equations to a system whose solutions are globally regular. The mapping from `old variables' $(t, \mbf{u}(\mbf{x},t))$ to `mapped
variables' $(\tau, \mbf{u}_{\mathrm{map}}(\mbf{x},\tau))$ is defined by:
 \begin{eqnarray}
\label{eq:mapped_time_and_velocity} \tau(t) \equiv \int_0^t||\mathbf{\bomega}(\cdot,t')||_{\infty}~dt'\,, \qquad
\mbf{u}_{\mathrm{map}}(\mbf{x},\tau)  \equiv \frac{\mbf{u}(\mbf{x},t)}{||\mathbf{\bomega}(\cdot,t)||_{\infty}}\,.
  \end{eqnarray}
The mapped time $\tau(t)$ is a strictly monotonically increasing function of $t$ and the mapped velocity is well defined
due to the following result:\\

\noindent \textbf{Lemma 1.} \emph{Assume the initial vorticity is not identically zero on $\Omega.$ Then the vorticity
supremum norm is positive: $||\mathbf{\bomega}(\cdot,t)||_{\infty} > 0 \,, \forall \,t \in [0,T).$} $\square$\\

\noindent \emph{Proof.} Using the well-known fact that the zeroes of vorticity are preserved by the flow for as long as the
vorticity is defined, we get: $||\mathbf{\bomega}(\cdot,t_1)||_{\infty} = 0$ for some $t_1 \in [0,T)$ if and only if
$||\mathbf{\bomega}(\cdot,t)||_{\infty} = 0 \,,\,\,\forall\,t \in [0,T).$ Since the initial vorticity is not identically
zero, the Lemma follows. $\square$

Moreover, the characteristics of each system are mapped accordingly.
 Let $\mbf{X}(\mbf{X}_0,t)$ and $\mbf{X}_{\mathrm{map}}(\mbf{X}_0,\tau)$ be the respective solutions of
 $$\frac{d}{dt}\mbf{X}(\mbf{X}_0,t) = \mbf{u}(\mbf{X}(\mbf{X}_0,t),t)\,,\qquad
 \frac{d}{d\tau}\mbf{X}_{\mathrm{map}}(\mbf{X}_0,\tau) = \mbf{u}_{\mathrm{map}}(\mbf{X}_{\mathrm{map}}(\mbf{X}_0,\tau),\tau)\,,$$
 with initial conditions $\mbf{X}_{\mathrm{map}}(\mbf{X}_0,0) = \mbf{X}(\mbf{X}_0,0) = \mbf{X}_0\,.$ Then, using equations
 \eref{eq:mapped_time_and_velocity} we get $\mbf{X}_{\mathrm{map}}(\mbf{X}_0,\tau(t)) =
 \mbf{X}(\mbf{X}_0,t)\,.$

The Euler equations \eref{eq:Euler} are equivalent to the system:
\begin{eqnarray}
\label{eq:mapped_Euler}
 \frac{\partial \mbf{u}_{\mathrm{map}}}{\partial \tau} + \mbf{u}_{\mathrm{map}} \cdot \nabla \mbf{u}_{\mathrm{map}} =
  - \nabla {p_{\mathrm{map}}} - \beta(\tau)\,\mbf{u}_{\mathrm{map}}\,, \qquad \quad \nabla \cdot \mbf{u}_{\mathrm{map}} = 0\,,
\end{eqnarray}
where all mapped fields are to be evaluated at $(\mbf{x},\tau)$ unless explicitly stated, and the coefficient $\beta(\tau)$
(damping if and only if $\int_{\Omega} |\mbf{u}_{\mathrm{map}}(\mbf{x},\tau)|^2~d^3x$ decreases in time) is defined by
\begin{eqnarray}
\label{eq:mapped_alpha_at_max} \beta(\tau) \equiv {\bomega_{\mathrm{map}}(\mbf{Y}(\tau),\tau) \cdot \left(\nabla
\mbf{u}_{\mathrm{map}}(\mbf{Y}(\tau),\tau)\right) \cdot \bomega_{\mathrm{map}}(\mbf{Y}(\tau),\tau)}\,,
\end{eqnarray}
where $\bomega_{\mathrm{map}}(\mbf{x},\tau)  \equiv  \nabla \times \mbf{u}_{\mathrm{map}}(\mbf{x},\tau)$ is the mapped
vorticity and $\mbf{Y}(\tau)$ is the position of mapped vorticity maximum, with
 $|{\bomega}_{\mathrm{map}}(\mbf{x},\tau)| \leq ||{\bomega}_{\mathrm{map}}(\cdot,\tau)||_\infty =
 |{\bomega}_{\mathrm{map}}(\mbf{Y}(\tau),\tau)| = 1\,\quad \forall \,\tau, \,\forall \,\mbf{x} \in \Omega \,.$

The existence of $\mbf{Y}(\tau)$ and $\beta(\tau)$ is a consequence of the assumed regularity of the
velocity field: $\mbf{u} \in C([0,T);H^s)\cap C^1([0,T);H^{s-1})\,,\,\,s\geq3\,,$ along with \mbox{Lemma 1} and a
well-known Sobolev lemma:
 \begin{eqnarray*}
||D^{p} \mbf{F}(\cdot)||_{\infty} \leq \widetilde{c}_s\,||\mbf{F}(\cdot)||_{H^{s+p}}\,,\,\forall\,p \in
\mathbb{Z}^+\cup\{0\},\, s>3/2\,,
 \end{eqnarray*}
 valid for 3D vector fields $\mbf{F}$ defined on $\Omega$, where $D^p$ is any combination of spatial derivatives of
 combined order $p$.

In order to be able to treat $\mbf{Y}(\tau)$ as a function, a selection mechanism is
required. Apart from mirror images --which do not pose any problem for the selection mechanism--, in the generic case
$\mbf{Y}(\tau)$ can be taken as a piecewise continuous function, with jumps at times when two competing local spatial
maxima of $|{\bomega}_{\mathrm{map}}(\mbf{x},\tau)|$, located at different places, coincide in value. The full treatment of
this generic case of discontinuous $\mbf{Y}(\tau)$ is possible and straightforward. But, for the sake of this
paper and to fix ideas, we assume from here on that $\mbf{Y}(\tau)$ is continuous. \mbox{We remark} that the main
results discussed in this paper (Theorem 2 and Corollary 3) can be extended with slight modifications to the general case of discontinuous $\mbf{Y}(\tau)$. It is worth mentioning that several initial conditions
considered as candidates for finite-time singularity \cite{Bra83,Ker93,HouLi06,Kid85} belong to the simplified case of
continuous $\mbf{Y}(\tau)$.

The new system \eref{eq:mapped_Euler}--\eref{eq:mapped_alpha_at_max} is globally regular:\\

 \noindent \textbf{Theorem 2.} \emph{The solution of the new system
(\ref{eq:mapped_Euler})--\eref{eq:mapped_alpha_at_max}
is regular for all finite values of the mapped time $0 \leq \tau < \infty$.} $\square$\\

\noindent \emph{Proof.} Let us suppose $\tau \in [0,\tau_0]$ with $0 < \tau_0 < \infty$ fixed. We obtain, using
equation (\ref{eq:mapped_time_and_velocity}),
 \begin{eqnarray}
 \nonumber
 \int_0^{t'} ||{\bomega}(\cdot,t'')||_\infty~dt''  \leq
\tau_0 < \infty\,,\quad \forall \, t' \in [0, t(\tau_0)]\,,
 \end{eqnarray}
  where $t(\tau)$ is the inverse map from new to original time.
According to BKM theorem \cite{BKM84}, this implies that in original variables all Sobolev norms of velocity
$||\mbf{u}(\cdot,t)||_{H^s}$ are bounded for all times $t \in [0,t(\tau_0)]\,.$ Also, by virtue of Lemma 1 the vorticity
modulus supremum norm is positive: $||{\bomega}(\cdot,t)||_\infty > 0, t \in [0,t(\tau_0)]\,.$ Therefore we can transform
to the mapped variables and use equation \eref{eq:Sobolev} to conclude that all Sobolev norms of mapped velocity
$||\mbf{u}_{\mathrm{map}}(\cdot,\tau)||_{H^s} = ||\mbf{u}(\cdot,t(\tau))||_{H^s}/||{\bomega}(\cdot,t(\tau))||_\infty$ are
bounded for $\tau \in [0, \tau_0]\,.$ Since $\tau_0$ is arbitrary the result $\mbf{u}_{\mathrm{map}} \in
C([0,\infty);H^s)\,,\,\,s\geq3$ is established. Along similar lines we can show that $\mbf{u}_{\mathrm{map}} \in
C^1([0,\infty);H^{s-1}).$ The proof is omitted here. $\square$\\

Theorem 2 holds even if the Euler system \eref{eq:Euler} has a finite-time singularity. This is very useful from a
computational point of view: one is sure that a careful numerical integration of the new system should give rise to a
regular solution, and this alone leads to a more reliable test of the validity of the finite-time singularity hypothesis.
This is in contrast with numerical simulations of the original Euler equations, where numerical errors and lack of
resolution may fail to find true finite-time singularities or may give rise to spurious ones, making any conclusions
unreliable.

\subsection{A robust criterion for finite-time singularity of Euler system} 
\label{sec:robust_criterion}
In the original variables there are
two types of constants of motion: the total kinetic energy of the fluid and the circulations of velocity field along
selected closed contours. For simplicity we focus on the \mbox{energy} \mbox{$E \equiv
\frac{1}{2}||\mbf{u}(\cdot,t)||_{L^2}^2 = \frac{1}{2}\int_{\Omega} |\mbf{u}(\mbf{x},t)|^2~d^3x,$} defined in terms of the
$L^2$-norm of velocity. In pseudospectral Euler simulations, constancy of energy is very robust and it holds even when
small-scale errors pervade the system. In a numerical simulation of the mapped system
\eref{eq:mapped_Euler}--\eref{eq:mapped_alpha_at_max} we do not have direct access to the Euler energy but we can define
the mapped energy:
 \begin{eqnarray}
 \label{eq:mapped_energy}
E_{\mathrm{map}}(\tau) \equiv \frac{1}{2}||\mbf{u}_{\mathrm{map}}(\cdot,\tau)||_{L^2}^2 = \frac{1}{2} \int_{\Omega}
|\mbf{u}_{\mathrm{map}}(\mbf{x},\tau)|^2~d^3x\,,
 \end{eqnarray}
which is not constant. From \eref{eq:mapped_time_and_velocity} we deduce that the supremum norm of the original vorticity
$||{\bomega}(\cdot,t(\tau))||_\infty$ can be obtained in terms of the Euler energy and mapped energy:
 \begin{eqnarray}
 \label{eq:omega_from_E}
 ||{\bomega}(\cdot,t(\tau))||_\infty  = \sqrt{\frac{E}{E_{\mathrm{map}}(\tau)}}\,,
 \end{eqnarray}
where the implicit inverse transformation $t(\tau)$ appears naturally. Notice that in practical applications, both the
original Euler energy $E$ and mapped energy $E_{\mathrm{map}}(\tau)$ are nonzero.

We reconstruct the original time $t(\tau)$ by integrating $dt/d\tau = 1/||{\bomega}(\cdot,t(\tau))||_\infty$ and using
\eref{eq:omega_from_E}:
 \begin{eqnarray}
 \label{eq:inverse_time}
  t(\tau)
= \int_0^\tau \frac{1}{||{\bomega}(\cdot,t(\tau'))||_\infty}~d\tau' = \int_0^\tau
\sqrt{\frac{E_{\mathrm{map}}(\tau')}{E}}~d\tau'\,.
\end{eqnarray}

This allows us to re-word the BKM criterion for singularity of Euler, in terms of the regular solution of the mapped
equations \eref{eq:mapped_Euler}--\eref{eq:mapped_alpha_at_max}. Let us define the quantity $t_{\mathrm{end}}$:
 \begin{eqnarray}
 \label{eq:t_end}
 t_{\mathrm{end}} \equiv \int_0^\infty \sqrt{\frac{E_{\mathrm{map}}(\tau)}{E}}~d\tau
 = \frac{\int_0^\infty ||\mbf{u}_{\mathrm{map}}(\cdot,\tau)||_{L^2}~d\tau}{||\mbf{u}(\cdot,0)||_{L^2}}\,.
 \end{eqnarray}

\noindent \textbf{Corollary 3.} \emph{The solution of the 3D Euler system \eref{eq:Euler} has a finite-time
singularity if and only if \,$t_{\mathrm{end}} <
\infty,$ in which case the singularity is attained at time $t = t_{\mathrm{end}}$ in the original variables.} $\square$\\

This represents a `duality' between the mapped regular system and the original system: a fast decay in time of the mapped
field is equivalent to a finite-time blowup of the original field.

\section{Proposed numerical simulations of mapped system and the finite-time blowup hypothesis of 3D Euler equations}
\label{sec:proposed}

\subsection{Advantages of the new approach} 

The numerical integration of the mapped system
\eref{eq:mapped_Euler}--\eref{eq:mapped_alpha_at_max} is completely independent of the Euler system \eref{eq:Euler}, and
its main advantages are:

\noindent \textbf{Adv.~1.} Since the profile of vorticity modulus is always bounded by 1, we expect a better regularity of
all fields involved in the computation of the next timestep data: velocity field and damping coefficient $\beta(\tau)$.
This will reduce numerical errors.

\noindent \textbf{Adv.~2.} As the profile of vorticity becomes more and more pronounced near its maximum (with vorticity
modulus equal to 1 at the maximum), the value of vorticity modulus away from the peak tends to zero. Accordingly, all
numerical noise generated away from the peak should be damped to zero. This is due to the damping produced by
$\beta(\tau)$, defined in equation \eref{eq:mapped_alpha_at_max}.

\noindent \textbf{Adv.~3.} Due to the boundedness of velocity, the Courant condition will produce a timestep which is
bounded from below (see Section \ref{subsec:numerics_3D_Euler} for details on the numerical code).

\noindent \textbf{Adv.~4.} The new approach puts in the same footing the two potential types of solutions of Euler
equations: the ones that are globally regular and the ones that blow up in a finite time. The reason being that for each of
the two types, the mapped version is regular. This gives a common platform to study the effect of initial conditions on the
singular character of the solutions.

\noindent \textbf{Adv.~5.} For solutions of Euler equations that are candidates for finite-time singularity, the
integration of the mapped regular equations will get further in time with better accuracy than the direct integration of
Euler equations. Once the new numerical codes to solve the mapped equations are validated, our method will serve as a
consistency test for the old Euler codes.

\subsection{Initial conditions} 
The guiding principle to set up initial conditions will be to use discrete
mirror symmetries, which enable us to reach high resolution by saving memory and computation time. Types of initial
conditions to be studied range from Taylor-Green-like to anti-parallel vortices, and we will apply to all these initial
conditions a growth-optimisation procedure which is under development.
 First we set up the initial conditions for the field $\mbf{u}(\mbf{x},t)$ at $t=0$: $\mbf{u}(\mbf{x},0) =
\mbf{u}_0(\mbf{x})$ such that $\nabla \cdot \mbf{u}_0(\mbf{x}) = 0 \,\,\forall \,\mbf{x}\in\Omega\,.$ We construct the
initial vorticity: $\bomega_0(\mbf{x}) \equiv \nabla \times \mbf{u}_0(\mbf{x})\,,$ and compute its supremum norm and the
position $\mbf{Y}(0)$ where it attains its maximum value:
 $||{\bomega}_0(\cdot)||_\infty \equiv \sup_{\mbf{x}\in\Omega} |{\bomega}_0(\mbf{x})| = |{\bomega}_0(\mbf{Y}(0))|\,.$
Finally, we compute the initial energy $E = \frac{1}{2} \int_{\Omega} |\mbf{u}_0(\mbf{x})|^2~d^3x\,.$

Following the mapping described in Section \ref{subsec:novel}, we now construct the initial conditions corresponding to the system
\eref{eq:mapped_Euler}--\eref{eq:mapped_alpha_at_max}. Notice that there is no need to explicitly `solve' for $\tau$ as a
function of $t$. We merely set $\tau(0) = 0\,.$ The initial conditions for the mapped velocity are thus
$\mbf{u}_{\mathrm{map}}(\mbf{x},0) = {\mbf{u}_0(\mbf{x})}/{||{\bomega}_0(\cdot)||_\infty}\,.$ The mapped vorticity
$\bomega_{\mathrm{map}} = \nabla \times \mbf{u}_{\mathrm{map}}$ satisfies initially $|\bomega_{\mathrm{map}}(\mbf{x},0)|
\leq 1 ,\,\forall \, \mbf{x} \in \Omega,$ and $|\bomega_{\mathrm{map}}(\mbf{Y}(0),0)| = 1\,.$

\subsection{Numerical integration of the mapped equations} 
\label{subsec:numerics_3D_Euler}
We will integrate numerically the mapped regular
system \eref{eq:mapped_Euler}--\eref{eq:mapped_alpha_at_max}. For this we will modify extensively two existing Euler codes,
in order to develop a new code where at each time step the following extra procedures are implemented:\\
%By virtue of this system, the mapped vorticity will satisfy $|\bomega_{\mathrm{map}}(\mbf{x},\tau)| \leq 1$ and
%$|\bomega_{\mathrm{map}}(\mbf{Y}(\tau),\tau)| = 1\,,$ where $\mbf{Y}(\tau)$ is the position of the maximum of mapped
%vorticity modulus.

\vspace{-.25cm} \noindent $\bullet$ an accurate modelling of the nonlocal damping coefficient $\beta(\tau)$, equation
\eref{eq:mapped_alpha_at_max}, using a high-order interpolation method to find the instantaneous position $\mbf{Y}(\tau)$
of the maximum of vorticity modulus and then computing the interpolated gradient of velocity at that point;\\

 \vspace{-.25cm} \noindent $\bullet$ a normalisation
algorithm that guarantees that the vorticity modulus be exactly equal to
 1 at $\mbf{x} = \mbf{Y}(\tau),$ as it should be analytically, and consequently $|\bomega_{\mathrm{map}}(\mbf{x},\tau)| \leq 1 \,\,\forall \mbf{x} \in \Omega$.\\

\vspace{-.25cm} The former procedure takes a reasonably small fraction of the computation time, depending on the required
interpolation accuracy. The latter procedure re-scales the fields uniformly in space, so it will increase only slightly the
computation time. The new code will be pseudospectral and parallel, using a Message-Passing-Interface protocol (MPI). Time
stepping will follow a high-order Runge-Kutta scheme, with a Courant condition to choose the value of the time step.

The following quantities need to be saved for post-processing analyses that can lead to conclusions about the finite-time
singularity hypothesis: damping coefficient \eref{eq:mapped_alpha_at_max}, mapped energy \eref{eq:mapped_energy} and mapped
circulations: $\sigma_{j}(\tau) = \oint_{{\mathcal{C}}_{j}} \mbf{u}_{\mathrm{map}}(\mbf{r},\tau) \cdot d \mbf{r}\,, \quad j
= 1, \ldots , N\,,$ where $\{{\mathcal{C}}_{j}\}_{j = 1}^N$ are selected fixed contours that depend on the type of mirror
symmetry used.

A reliability time $\tau_{\mathrm{rel}}$ will be determined for each numerical simulation performed, so that the simulation
is reliable for times $\tau<\tau_{\mathrm{rel}}$. Reliability checks to be produced are: conservation of quantities $K_{j}
= \sigma_{j}(\tau)/[E_{\mathrm{map}}(\tau)]^{1/2}$ defined in terms of mapped energy and circulation, classical resolution studies and analyticity-strip convergence studies \cite{Sul83}.
  
\subsection{Searching for late-time trends of singular/nonsingular behaviour}
 We proceed to interpolate the
saved quantities as functions of time $\tau$ between $0$ and $\tau_{\mathrm{rel}},$ in particular the supremum norm of the
original vorticity $||{\bomega}(\cdot,t(\tau))||_\infty = \left({E}/{E_{\mathrm{map}}(\tau)}\right)^{1/2}\,.$

Notice that in the usual setting of Euler equations, researchers have tried various types of fits for the late-time trend
of the vorticity supremum norm as a function of time $t$. For example, a double exponential
\mbox{$||{\bomega}(\cdot,t)||_\infty \approx a\,\exp (\exp (b\,t))$} in \cite{HouLi06} and a power law
\mbox{$||{\bomega}(\cdot,t)||_\infty \approx c \,(T_*-t)^{-d}$} in \cite{BusKerr08}. We can classify all types of fits in
two classes: in one class, the fits that \emph{do not} represent a finite-time singularity and in the other class, the fits
that represent \emph{explicitly} a finite-time singularity. In both classes, $||{\bomega}(\cdot,t)||_\infty$ increases
monotonically with time $t$.

In contrast, in the new setting in terms of time $\tau$ and mapped variables, the frontier between the two classes is more
subtle. According to Theorem 2, the mapped fields are regular for all times $\tau < \infty.$ In the two classes, any fit
for $\tau\to\infty$ trends of $||{\bomega}(\cdot,t(\tau))||_\infty$ must be continuous.

In the new setting, there is a more robust means to conclude on the finite-time singularity hypothesis. From Corollary 3,
the blowup time is $t_{\mathrm{end}} = \int_0^\infty ||{\bomega}(\cdot,t(\tau'))||_\infty^{-1}~d\tau'  = \int_0^\infty
\left({E_{\mathrm{map}}(\tau')}/{E}\right)^{1/2}~d\tau'\,.$ Therefore, a late-$\tau$ fit for
$E_{\mathrm{map}}(\tau)$ such that this integral converges, provides robust and reliable evidence in favour of
the finite-time singularity hypothesis. Conversely, if the integral diverges, the evidence will be against the hypothesis.

To illustrate these ideas, let us suppose for example that we have obtained a late-time trend
 $||{\bomega}(\cdot,t(\tau))||_\infty \approx K\,\tau^\gamma\,,$
where $K>0$ and $\gamma > 0$ are fit parameters. Using equation \eref{eq:inverse_time} we solve for $t(\tau)$ and,
subsequently, find $||{\bomega}(\cdot,t)||_\infty$ by inversion and substitution. We get:
 \begin{eqnarray}
 \nonumber
 t \approx \left\{
 \begin{array}{rl}
 \frac{1}{(1-\gamma)\,K} \,\tau^{1-\gamma} + t_0\,, & \gamma \neq 1\\
 \frac{1}{K} \ln \tau + t_0\,, &\gamma = 1
 \end{array}
 \right.
 \qquad  \quad
 ||{\bomega}(\cdot,t)||_\infty \approx \left\{
 \begin{array}{rl}
 \widetilde{K} \left[(1-\gamma)\left(t-t_0\right)\right]^{\frac{\gamma}{1-\gamma}}\,, & \gamma \neq 1\\
 K \exp K\left(t-t_0\right)\,, &\gamma = 1
 \end{array}
 \right.
 \end{eqnarray}
 where $t_0$ is a constant of integration and $\widetilde{K} > 0$ is a constant. We see that the simple power-law fit $||{\bomega}(\cdot,t(\tau))||_\infty \approx
K\,\tau^\gamma\,,$ leads to a prediction of finite-time singularity if and only if the exponent $\gamma > 1\,,$ which is in
complete agreement with Corollary 3. In the case $\gamma>1,$ the hypothetical singularity time comes directly from equation
\eref{eq:t_end}, and is not a fit parameter.\\

\section{Other equations that admit a mapping to a regular system: the case of ideal magnetohydrodynamics}
\label{sec:other}

The analysis in the previous two Sections can be extended to include other equations of physical and mathematical interest such as 3D Navier-Stokes, ideal 2D and 3D magnetohydrodynamics (MHD) and even 1D inviscid Burgers (see Section \ref{sec:Burgers}). The basic ingredient needed is a BKM-type of theorem. For simplicity we discuss briefly the method for the equations of ideal 2D MHD:
\begin{equation}
\label{eq:MHD}
 \left(\partial_t + \mbf{u} \cdot \nabla \right) \omega = \mbf{b} \cdot \nabla j\,, \qquad \left(\partial_t + \mbf{u} \cdot \nabla \right) \psi = 0\,,
\end{equation}
where $\nabla$ denotes the 2D gradient, the vector fields $\mbf{b}$ (magnetic induction) and $\mbf{u}$ (velocity) are defined in terms of the scalars $\psi$ (magnetic potential) and $\omega$ (vorticity) by $\mbf{b} = \nabla^\perp \psi\,, \quad \mbf{u} = \nabla^\perp \triangle^{-1} \omega\,,$ and the current $j$ is given by $j = \triangle \psi.$ The perpendicular gradient is defined by $\nabla^\perp \equiv \mbf{\hat{k}} \times \nabla\,,$ where $\mbf{\hat{k}}$ denotes the out-of-plane unit vector.

A theorem of BKM type, valid for 2D and 3D MHD, was found by Caflisch, Klapper and Steele \cite{Caf97} and it states, for the 2D case:\\

\noindent \textbf{BKM theorem for MHD.} (Caflisch et al.)  \emph{The solution of the MHD equations \eref{eq:MHD} is regular up to and including time $t$ if and only if the following integral is bounded:
\begin{equation}
\label{eq:mapped_time_MHD}
 \tau(t) = \int_0^t \left(\|\omega(\cdot,t')\|_\infty + \|j(\cdot,t')\|_\infty\right)~dt'\,. ~ \square
\end{equation}}

Corresponding to this time transformation we define mapped fields as follows:
\begin{equation} 
\label{eq:mapped_fields_MHD}
\Omega(\mbf{x},\tau) = \frac{\omega(x,t)}{\|\omega(\cdot,t)\|_\infty + \|j(\cdot,t)\|_\infty}\,, \quad 
 \Psi(\mbf{x},\tau) = \frac{\psi(x,t)}{\|\omega(\cdot,t)\|_\infty + \|j(\cdot,t)\|_\infty}\,,
\end{equation}
along with the mapped velocity $\mbf{U} = \nabla^\perp \triangle^{-1} \Omega$ and mapped current $J = \triangle \Psi\,.$

In the new variables we have the identity $\|\Omega(\cdot,\tau)\|_{\infty} + \|J(\cdot,\tau)\|_{\infty} = 1\quad \forall \, \tau\,,$ which assures the boundedness of the supremum norms of mapped vorticity and current. It will be useful to define the respective spatial positions $\mbf{Y}_1(\tau), \mbf{Y}_2(\tau)$ where the mapped vorticity $\Omega$ and current $J$ attain their respective maxima. We write
\begin{eqnarray}
 \|\Omega(\cdot,\tau)\|_{\infty} = |\Omega(\mbf{Y}_1(\tau),\tau)|\,,\quad \nabla \Omega(\mbf{Y}_1(\tau),\tau) = 0\,,\\
 \|J(\cdot,\tau)\|_{\infty} = |J(\mbf{Y}_2(\tau),\tau)|\,, \quad \nabla J(\mbf{Y}_2(\tau),\tau) = 0\,.
\end{eqnarray}
As discussed in the 3D Euler case for the position of maximum vorticity, the positions $\mbf{Y}_1(\tau), \mbf{Y}_2(\tau)$ have discontinuities when competing maxima of vorticity and current coincide. However, in a fast-growth scenario, we expect these discontinuities to be distributed as a finite number of isolated points in the time axis, so after a transient we can consider these functions as continuous. The discontinuous case can be treated without difficulty but for simplicity of presentation we assume that these functions are continuous.

After the bijective mapping \eref{eq:mapped_time_MHD}--\eref{eq:mapped_fields_MHD}, the new variables satisfy the following system of equations:
\begin{equation}
 \label{eq:mapped_MHD}
 \left(\partial_\tau + \mbf{U} \cdot \nabla \right) \Omega = \mbf{B} \cdot \nabla J - \gamma(\tau)\,\Omega\,, \qquad \left(\partial_\tau + \mbf{U} \cdot \nabla \right) \Psi = -\gamma(\tau) \, \Psi\,,
\end{equation}
where the damping/amplifying coefficient $\gamma(\tau)$ is a quadratic function of the new variables. Explicitly, we have
\begin{equation}
\label{eq:gamma}
 \gamma(\tau) = \left[ \mathrm{sgn} \Omega\,\mbf{B} \cdot \nabla J \right]_1 + \left[ \mathrm{sgn} J\,\mbf{B} \cdot \nabla \Omega \right]_2 + \left[2\, \varepsilon_{p q} \partial_r B_p\,\partial_r U_q \right]_2\,,
\end{equation}
where $\mathrm{sgn}$ represents the \emph{signum} function, $\varepsilon_{p q}$ are the components of the 2-dimensional fully antisymmetric tensor with $\varepsilon_{12} =1,$ Einstein sum convention over repeated indices is assumed, and 
$ [F(\mbf{x},\tau)]_1 \equiv F(\mbf{Y}_1(\tau),\tau)\,,\quad  [F(\mbf{x},\tau)]_2 \equiv F(\mbf{Y}_2(\tau),\tau)\,,
$ for any function $F(\mbf{x},\tau)\,.$ The coefficient $\gamma(\tau)$ will be damping if and only if $\int (|\mbf{U}(\mbf{x},\tau)|^2 + |\mbf{B}(\mbf{x},\tau)|^2 )~d^2 x$ grows in time. 

In a way completely analogous to the 3D Euler case we can prove the global regularity of the MHD mapped system:\\

\noindent \textbf{Theorem 4.} \emph{Assuming smooth initial conditions, the solutions to the mapped system \eref{eq:mapped_MHD}--\eref{eq:gamma} is regular for all times $\tau>0. \,\square$}\\

Also, it is possible to reword the BKM-type of theorem for MHD in terms of the mapped fields. Defining the original energy (constant) and the mapped energy (not constant) by
\begin{equation}
{\mathcal E} \equiv \frac{1}{2}\,\int (|\mbf{u}(\mbf{x},t)|^2 + |\mbf{b}(\mbf{x},t)|^2 )~d^2 x\,,\quad {\mathcal E}_{\mathrm{map}}(\tau) \equiv \frac{1}{2}\,\int (|\mbf{U}(\mbf{x},\tau)|^2 + |\mbf{B}(\mbf{x},\tau)|^2 )~d^2 x\,,
\end{equation}
we obtain $\|\omega(\cdot,t(\tau))\|_\infty + \|j(\cdot,t(\tau))\|_\infty = \sqrt{{\mathcal E}/{\mathcal E}_{\mathrm{map}}(\tau)}$, where $t(\tau)$ denotes the inverse of transformation \eref{eq:mapped_time_MHD}. Defining the singularity time $t_{\mathrm{end}} = \int_0^\infty \sqrt{{\mathcal E}_{\mathrm{map}}(\tau)/{\mathcal E}}~d\tau\,,$ we have:\\

\noindent \textbf{Corollary 5.} \emph{The solution of the MHD system \eref{eq:MHD} has a finite-time
singularity if and only if $t_{\mathrm{end}}  < \infty,$ in which case the singularity is attained at time $t = t_{\mathrm{end}}$ in the original variables.} $\square$\\

Notice that the amplifying coefficient is generically quadratic in the mapped MHD fields, in contrast to the Euler case where the coefficient is linear in the fields (for inviscid Burgers, see Section \ref{sec:Burgers}, the coefficient turns out to be a constant!). This extra nonlinearity might determine an extra dealiasing factor in numerical applications, but this is to be established in a forthcoming paper.

\section{Proof of concept: an illustrative numerical example}
\label{sec:Burgers}
\subsection{The inviscid Burgers equation in one dimension}
To illustrate the usefulness of the mapping \eref{eq:mapped_time_and_velocity}, we have produced numerical simulations of a well-known nonlinear wave equation, called ``inviscid Burgers equation'' in modern literature, although it was introduced and solved at least as early as 1808 by Poisson \cite{Poisson1808}.

We have chosen this equation for illustration purposes for three reasons: (i) its solutions are known to blow up in a finite time, (ii) it posseses a BKM type of criterion for blowup, and (iii) all relevant norms (supremum of gradient, Sobolev) can be found analytically in terms of the initial conditions, by the method of characteristics. Point (i) ensures that the example is nontrivial because the numerical method is challenging. Point (ii) allows us to apply the time mapping in complete analogy to the case of 3D Euler equations. Point (iii) allows us to use the analytical expressions for the relevant norms, to produce reliable comparisons and validation tests of numerical datasets from both original equation and mapped regular system.

We consider a spatially periodic setting for a scalar field $u(x,t) \in \mathbb{R}$ depending on $x \in \mathbb{R}$ and $t \in [0,t_{\mathrm{end}})$, such that $u(x+2\pi,t) = u(x,t)$, with initial condition $u(x,0) = u_0(x)$ also periodic and of class ${C}^{\infty}$. The field $u(x,t)$ satisfies the differential equation
\begin{equation}
\label{eq:Burgers}
 u_t + u \,u_x = 0\,,
\end{equation}
where the subscript denotes differentiation with respect to the corresponding argument. The blowup time $t_{\mathrm{end}} > 0$ is given in terms of the initial condition $u_0(x)$ through the well-known formula $t_{\mathrm{end}} = - (\min_{x_0 \in [0,2\pi]} u_0'(x_0))^{-1}$. Relevant constants of motion in this case are the energy $E \equiv \frac{1}{2} \int_0^{2\pi}|u({x},t)|^2~dx$ and `circulation' $\sigma_0 \equiv  \int_0^{2\pi}u({x},t)~dx\,.$

The analogue of supremum norm of vorticity in 3D Euler equations is for inviscid Burgers the supremum norm of $u_x,$ defined by
\begin{equation}
\label{eq:supnorm_Burgers} \left\|u_x(\cdot,t)\right\|_{\infty} \equiv \sup_{x \in [0,2\pi]} \left|u_x(x,t)\right|\,.  
\end{equation}
For the sake of simplicity of presentation we avoid the situation of competing local maxima of the gradients, by assuming that $\left\|u_0'(\cdot)\right\|_{\infty} =  |\min_{x_0 \in [0,2\pi]} u_0'(x_0)|\,.$ Then, there is a well-known analytical expression for the norm \eref{eq:supnorm_Burgers}, obtained by solving equation \eref{eq:Burgers} along the characteristics:
\begin{equation}
\label{eq:supnorm_example_Burgers}
 \left\|u_x(\cdot,t)\right\|_{\infty}  = \frac{1}{t_{\mathrm{end}} - t}\,.
\end{equation}
For times $t < t_{\mathrm{end}}$, it can be shown by explicit computations that the solution remains in the class ${C}^{\infty}$, while for $t= t_{\mathrm{end}}$ all supremum norms of the spatial derivatives diverge. By direct inspection it is therefore evident that, for inviscid Burgers, the analogue of 3D Euler's BKM theorem is:\\

\noindent \textbf{Theorem 6.} The solution $u(x,t)$ of equation \eref{eq:Burgers} is bounded up to and including time $t$ if and only if the following integral is bounded:
\begin{equation}
\label{eq:mapped_time_Burgers}
 \tau(t) = \int_0^t \left\|u_x(\cdot,t') \right\|_{\infty}~dt' \,. ~ \square 
\end{equation}

The above equation defines the mapped time $\tau(t)$, and accordingly we define the mapped velocity field $v(x,\tau) = u(x,t)/\|u_x(\cdot,t)\|_{\infty}\,.$ The bijectively-mapped equation for $v(x,\tau)$ is
\begin{equation}
\label{eq:mapped_Burgers}
   v_{\tau} + v \, v_{x} = -\alpha(\tau)\, v\,,
\end{equation}
where the damping/amplifying coefficient is defined by $\alpha(\tau) \equiv - v_{x}(Y(\tau),\tau),$ and $Y(\tau)$ is the spatial position of the maximum of $\left|v_{x}(x,\tau)\right|\,.$ By virtue of the above transformation we have $|\alpha(\tau)| = \left|v_{x}(Y(\tau),\tau)\right| =  \left\|v_{x}(\cdot,\tau)\right\|_{\infty} = 1\,.$ Therefore, in our mapped equation \eref{eq:mapped_Burgers}, the damping/amplifying coefficient can generally take only two values: either $\alpha(\tau) = 1$ or $\alpha(\tau) = -1.$ Discontinuities of this coefficient can arise when two local maxima of $\left|v_{x}(x,\tau)\right|$ compete. To avoid this situation, we have chosen for simplicity the initial condition such that $\left\|u_0'(\cdot)\right\|_{\infty} = |\min_{x_0 \in [0,2\pi]} u_0'(x_0)|\,.$ So, in our example the damping coefficient turns out to be a constant, equal to unity.

\subsection{Mapping fields} 
From equations \eref{eq:supnorm_example_Burgers}, \eref{eq:mapped_time_Burgers} we obtain
\begin{equation}
 \label{eq:explicit_time_map}
 \tau(t) = - \ln (t_{\mathrm{end}} - t), \quad  \left\|u_x(\cdot,t(\tau))\right\|_{\infty} = \frac{1}{t_{\mathrm{end}}}\, \exp \tau\,.
\end{equation}
In numerical simulations of the mapped equation \eref{eq:mapped_Burgers} we have direct access to the mapped energy and circulation
\begin{equation*}
 E_{\mathrm{map}}(\tau) \equiv \frac{1}{2} \int_0^{2\pi}
|v({x},\tau)|^2~dx\,, \quad  \sigma(\tau) \equiv  \int_0^{2\pi}
v({x},\tau)~dx\,.
\end{equation*}
In a manner completely analogous to what was done in \sref{sec:robust_criterion}, we obtain the identities
$ \left\|u_x(\cdot,t(\tau))\right\|_{\infty} = \sqrt{\frac{E}{E_{\mathrm{map}}(\tau)}}\,, \,\,\,\,  t_{\mathrm{end}} = \int_0^\infty \sqrt{\frac{E_{\mathrm{map}}(\tau)}{E}}~d\tau\,,$
and the constant of motion
$K = \frac{\sigma(\tau)}{\sqrt{E_{\mathrm{map}}(\tau)}} = \frac{\sigma_0}{E}\,.$ 

Correspondingly, the mapped energy and circulations can be obtained explicitly:
 \begin{equation}
\label{eq:explicit_mapped_energy}
 E_{\mathrm{map}}(\tau) = (t_{\mathrm{end}})^2 \,E\, \exp(-2\tau)\,,\quad \sigma(\tau) = \sigma_0 \,t_{\mathrm{end}}\,\exp(-\tau)\,.
\end{equation}

\subsection{Numerical simulations of inviscid Burgers and mapped system: comparison of data with the analytical solution} 

For simplicity we will take the initial condition $u_0(x_0) = 1 + \cos x_0\,,$ which satisfies the assumptions at the beginning of this Section. The blowup time is $t_{\mathrm{end}} = 1.$ The constants of motion for inviscid Burgers studied here, energy and circulation, are respectively $E = \frac{1}{2} \int_0^{2\pi}|u_0({x})|^2~dx = 1.5 \pi \,, \quad \sigma_0 = \int_0^{2\pi} u_0({x})~dx = 2\pi\,.$  We have put deliberately an additive constant in this initial condition so that the position of maximum of $|u_x(x,t)|$ moves in time, which is a necessary test for the algorithm that searches for the accurate position of the maximum in the numerical solution of the mapped equations.

In order to produce a proper comparison of the data arising from the numerical simulation of the inviscid Burgers equation \eref{eq:Burgers} and the mapped equation \eref{eq:mapped_Burgers}, 
we approach the problem using the following principles:\\

\vspace{-.25cm} \noindent $\bullet$ The basic numerical method used for both systems should be identical. We choose as a method the classical pseudospectral with leap-frog time advancement (second order accurate), which uses a constant time stepping determined by a CFL condition with a typical Courant number between 1/5 and 1/20. This method has been used previously in numerical simulations of inviscid Burgers \cite{Sul83}. Our method differs slightly with the one used in the cited reference, in that we do not mix odd- and even-timestep data every 20 steps. We found that it is not needed to do this mixing to obtain convergence. The damping term in the regular system is treated using a Crank-Nicholson scheme, which keeps the second-order accuracy.\\

\vspace{-.25cm} \noindent $\bullet$ Unlike the integration of the inviscid Burgers equation, the integration of the regular system requires a normalisation of the fields every time step, so that the supremum norm is always equal to one. For this, at each time step the \emph{spatial position} of the maximum of $|v_x(x,\tau)|$ is located using an accurate method of order 16, then the function $|v_x(x,\tau)|$ is interpolated at that position to obtain $\|v_x(\cdot,\tau)\|_{\infty}$, which is used to rescale the field $v(x,t)$. This procedure takes only a small fraction (less than 10 percent) of the full computer time used in a time stepping. To compensate for this extra time, we have proportionally reduced the CFL number in the integration of inviscid Burgers equation when producing comparisons, so that we will always be comparing runs with the same spatial resolution, that took approximately the same computer time and number of timestep iterations.\\

\begin{figure}[h]
 \includegraphics[width=7.5cm,height=4.5cm,angle=0]{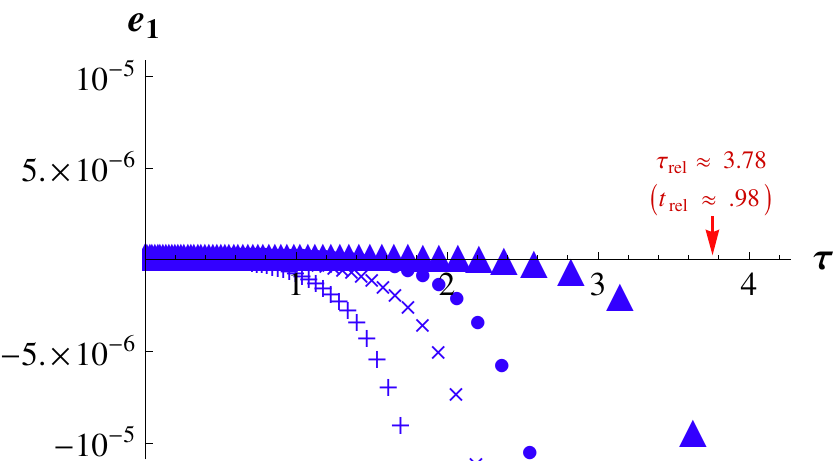}
\hspace{1cm}
 \includegraphics[width=7.5cm,height=4.5cm,angle=0]{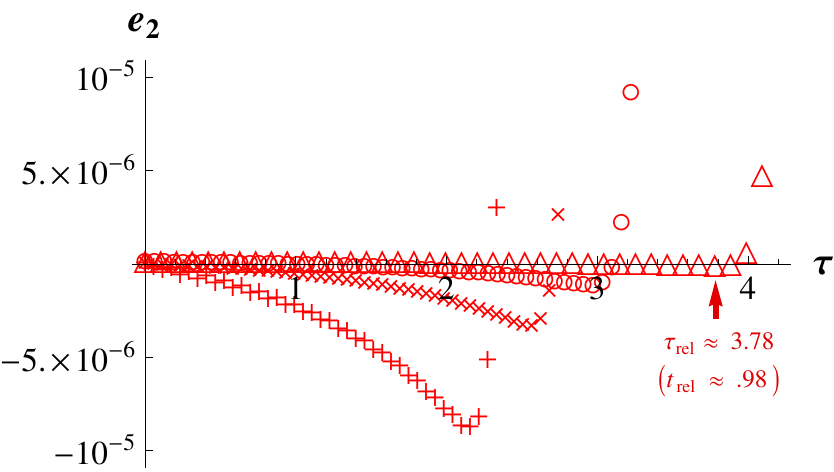}
\caption{\label{fig:res_study} Colour online. \textbf{Left panel:} Resolution study of data from integration of inviscid Burgers equation. Plots, for different spatial resolutions, of the error $e_1$ with respect to the analytical solution. Symbols (blue online): `$+$', `$\times$', `$\bullet$' and `$\blacktriangle$' correspond respectively to resolutions 1024, 2048, 4096 and 16384. For reference, the reliability time computed from data of integration of \emph{mapped system} at resolution 16384 is indicated by an arrow (orange online). \textbf{Right panel:} Resolution study of data from integration of regular mapped system. Plots, for different spatial resolutions, of the error $e_2$ with respect to the analytical solution. Symbols (red online): `$+$', `$\times$', `$\circ$' and `$\triangle$' correspond respectively to resolutions 1024, 2048, 4096 and 16384.}
\end{figure}

\vspace{-.25cm} We will compare the numerical data with the analytical solution. We map the data for the maximum vorticity obtained from the integration of inviscid Burgers equation \eref{eq:Burgers}, using the map $\tau = - \ln(1-t)$, so that in all plots below we have $\tau$ as the time variable. From equation \eref{eq:explicit_time_map} we should have $ \left\|u_x(\cdot,t(\tau))\right\|_{\infty}  = \exp \tau.$ Therefore we plot the error $e_{1}(\tau) = 2(\tau - \ln \left\|u_x(\cdot,t(\tau))\right\|_{\infty})$ versus $\tau$, where the data used comes from the integration of inviscid Burgers equation. If the numerical data is sensible, this error should be close to zero. Figure \ref{fig:res_study}, left panel, shows a resolution study with resolutions $N = 1024, 2048, 4096$ and $16384\,.$

For comparison, we use the data for the mapped energy $E_{\mathrm{map}}(\tau),$ obtained from the independent integration of the mapped system \eref{eq:mapped_Burgers}. From equation \eref{eq:explicit_mapped_energy} we should have $E_{\mathrm{map}}(\tau) = 1.5\, \pi\, \exp(-2\tau).$ We plot, versus $\tau$, the error $e_{2}(\tau) = 2 \tau + \ln E_{\mathrm{map}}(\tau) - \ln (1.5\, \pi)$ which analytically should be equal to $e_{1}(\tau)$ and therefore close to zero. Figure \ref{fig:res_study}, right panel, shows a resolution study with resolutions $N = 1024, 2048, 4096$ and $16384.$ 

\begin{figure}[h]
 \includegraphics[width=7.5cm,height=4.5cm,angle=0]{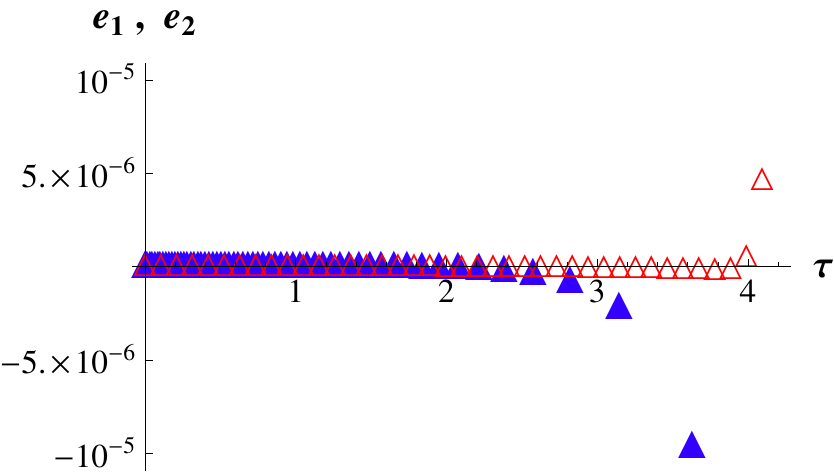}
\hspace{1cm}
 \includegraphics[width=7.5cm,height=4.5cm,angle=0]{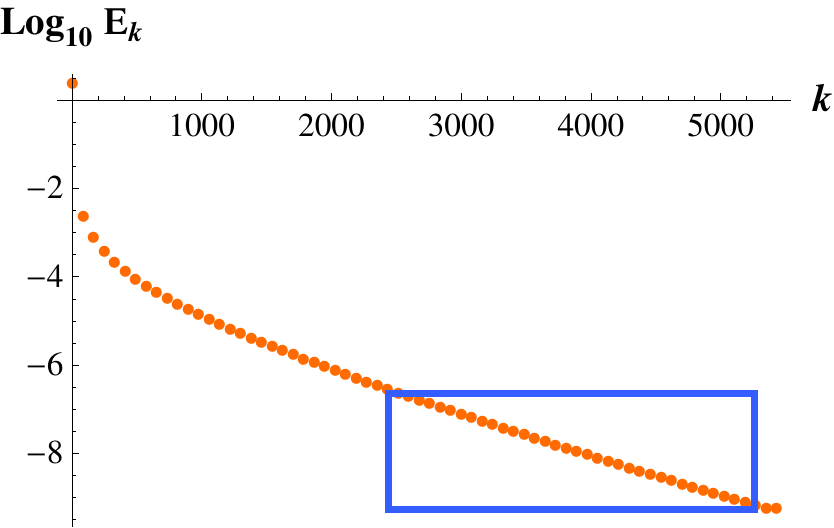}
\caption{\label{fig:accuracy} Colour online. \textbf{Left Panel:} Comparison of errors with respect to the analytical solution, of datasets obtained from integration of inviscid Burgers equation (error $e_1$, blue bullets) and regular mapped system (error $e_2$, red open circles), at fixed resolution 16384. \textbf{Right panel:} Log-lin plot of energy spectrum $E_k \propto |\hat{v}(k,t)|^2$ (orange bullets), of the solution $v(x,t)$ of the regular mapped equations at time $\tau = 4.2,$ using data from numerical integration of regular mapped system. The resolution is 16384, corresponding to  $k_{\max} = 5461$ by the 2/3-rd dealiasing rule. It is possible to observe at least two decades of exponential decay of the spectrum (rectangular region, light blue online), in agreement with the criterion of reliability commonly used in the analiticity-strip method.}
\end{figure}

Notice that the scale used for both figures is the same. We observe that the convergence towards the analytical solution, as one increases the resolution, is much faster for the mapped system data than for the original inviscid Burgers system. Also, it is evident that the numerical simulation of the mapped system captures the singularity with better accuracy in the time domain for a given spatial resolution: figure \ref{fig:accuracy}, left panel, shows a comparison of $N=16384$ data from both methods. The error of the numerical data of the mapped system is one order of magnitude less than the error of the numerical data of the original system. This effect is due in part to the fact that the integration of the original inviscid Burgers system uses a constant timestep and therefore is bound to misrepresent the solution at times that are close to the singularity time $t_{\mathrm{end}}=1$. In contrast, the integration of the mapped system also uses a constant timestep but the singularity is at $\tau_{\mathrm{end}}=\infty\,,$ so the solution is well-resolved as long as the spatial resolution is not lost due to the localisation of the structures.

Finally, notice that for a given spatial resolution both original system and regular system have a `reliability time' $\tau_{\mathrm{rel}}$ after which the fields are not well-resolved spatially anymore. One way to quantify the reliability time is by using the analiticity-strip method \cite{Sul83}, see also \cite{Bra83,BusKerr08} where the method is applied to 3D Euler equations. At resolution 16384, a calculation using the analiticity-strip method would give $\tau_{\mathrm{rel}} \approx 4.2,$ corresponding to original time $t_{\mathrm{rel}} \approx .99$ (see figure \ref{fig:accuracy}, right panel, and its caption). An ad-hoc and more conservative way to visualise the reliability time is to look at figure \ref{fig:res_study}, right panel. The reliability time for a given spatial resolution is approximately the time when the local minimum of the error occurs. For example, at resolution 16384 the reliability time is $\tau_{\mathrm{rel}} \approx 3.78,$ corresponding to original time $t_{\mathrm{rel}} \approx 0.98\,.$ It is clear from figure \ref{fig:accuracy}, left panel, that the data from the mapped system agrees with the analytical solution up to times very close to the corresponding reliability time, while the data for inviscid Burgers system begins to diverge earlier.

As an extra check, using data from the numerical solution of the mapped regular system, we have confirmed the constancy of $K = \sigma(\tau)/\left(E_{\mathrm{map}}(\tau)\right)^{1/2} = \sigma_0/E^{1/2} = 2 \left(\pi/3\right)^{1/2}$ for all runs, within one part in $10^{-7}$ for resolution 1024, and well beyond the reliability time.

\section{Conclusion and Discussion}
\label{sec:concl}

We have established, for several equations of physical interest, a nonlinear time transformation and field rescaling that is based on a BKM-type of theorem, and which maps bijectively the original equations to a system that is globally regular in time.

We have learnt from the simple example of inviscid Burgers that the numerical integration of a mapped regular system produces more accurate and reliable results than the integration of the original system. The only way that the integration of the original system can match the regular system, is by upgrading the numerical method to an adaptive time stepping, based on the transformation \eref{eq:mapped_time_Burgers}, so that $d\tau = dt \times \|u_x(\cdot,t)\|_{\infty}$ is kept constant. In addition, we validated a procedure of spatial interpolation (of order 16) used to compute accurately the value of the supremum norm $\|v_x(\cdot,\tau)\|_{\infty}$. This procedure is essential for the numerical integration of the corresponding mapped regular system, not only in inviscid Burgers but also in 3D Euler, Navier Stokes and magnetohydrodynamics. It is worth mentioning that we also used this interpolation procedure to compute $\|u_x(\cdot,t)\|_{\infty}$ from the numerical data of inviscid Burgers equations. If we had just computed the collocation-point maximum value of the same quantity, we would have obtained a prohibitive amount of spurious oscillations (figure not shown).

In a forthcoming paper the method will be implemented to integrate the 3D Euler equations. We expect the integration of the regular mapped system to produce a significant improvement in accuracy. However in this case no analytical solution is at hand, so in order to draw robust conclusions on the trends we will need to make a late-time fit along with a careful determination of the reliability time. We are considering the possibility to include also a time-dependent spatial deformation in order to capture the localised structures as they develop. This procedure is in the spirit of \cite{McLau86,Land91} and the recent review \cite{Budd10}.

\section{Acknowledgements}

The author thanks E. Cox, F. Dias, D. D. Holm, R. M. Kerr,  L. \'O N\'araigh, R. Temam, E. S. Titi and an anonymous referee for useful comments. Support for this work was provided by UCD Seed Funding project SF304 and IRCSET Ulysses project ``Singularities in three-dimensional Euler equations: simulations and geometry''.

\bibliographystyle{unsrt}
\bibliography{bibli_Euler_Regular_Final}

\end{document}